\documentclass[useAMS,usenatbib,fleqn]{mn2e}  
\usepackage{mathptmx}
\usepackage{graphicx}
\usepackage{amsmath}
\usepackage{amssymb}
\usepackage{psfig}
\usepackage{multirow}
\usepackage{cases}
\usepackage{subfigure}
\def\Msun{\hbox{$\thinspace M_{\odot}$}}

\def\xmm{{\it XMM-Newton}}

\def\Mbh{\hbox{$M_{\rmn{BH}}$}}
\def\xmm{{\it XMM-Newton}}
\newcommand{\pg}{PG 1244+026~\/}

\def\gsim{\mathrel{\hbox{\rlap{\hbox{\lower4pt\hbox{$\sim$}}}\hbox{$>$}}}}
\def\lsim{\mathrel{\hbox{\rlap{\hbox{\lower4pt\hbox{$\sim$}}}\hbox{$<$}}}}

   \title[X-ray time delays in PG 1244+026]{X-ray time delays in the Narrow Line Seyfert 1 galaxy PG 1244+026}
   \author[W. N. Alston et al.]{W. N. Alston$^{1,2}$\thanks{wna@ast.cam.ac.uk},
        C. Done$^{3}$, S. Vaughan$^{1}$\\
        $^{1}$University of Leicester, X-ray \& Observational Astronomy Group, Department of Physics and Astronomy, Leicester, LE1 7RH. \\
        $^{2}$Institute of Astronomy, Madingley Rd, Cambridge, CB3 0HA. \\
        $^{3}$Department of Physics, University of Durham, South Road, Durham, DH1 3LE. 
}

\date{Accepted 2014 January 1.  Received 2013 December 16; in original form 2013 October 20}
 
\pagerange{\pageref{firstpage}--\pageref{lastpage}} \pubyear{2013}

\begin{document}
\label{firstpage}
\maketitle

\begin{abstract}
We analyse the X-ray time-lags in the Narrow Line Seyfert 1 (NLS1)
galaxy PG 1244+026 ($M_{\rm BH}\sim 10^7 \Msun$, $L/L_{\rm Edd}\sim 1$).
The time delay between the soft (0.3--0.7~keV) and harder (1.2--4.0~keV)
variations shows shows the well established switch from hard lags at
low frequencies to soft lags at high frequencies.  The low frequency
hard lags are qualitatively consistent with the propagation of fluctuations model, with some long-timescale response of the reflection component.
The high frequency soft lag appears to extend over a wide frequency
band, that we divide this into two narrow frequency ranges, and examine
the lag as a function of energy for each of these.  At high frequencies
the soft excess emission is delayed with respect to the harder energy
bands, without any corresponding strong, hard X-ray reflection signature.  At even higher
frequencies a soft lag is seen at the softest energies, as well as
tentative evidence for an iron K$\alpha$ reverberation signal. 
These results point to the importance of reprocessing as well as
reflection in determining the lags in NLS1s. 
\end{abstract}


\maketitle
%

\section{Introduction}
\label{sect:intro}

The X-ray spectra of AGN usually show a soft X-ray excess (hereafter, soft excess) above
the low energy extrapolation of the 2-10~keV power law emission.  The
shape of this soft excess is often equally well modelled by either
smeared, ionised reflection (presumably from the surface of the
putative accretion disc: \citealt{crummy06}; \citealt{walton13}) or by a
separate soft continuum component
(\citealt{vaughan02}; \citealt{GeirlinksiDone04}). X-ray spectral
fitting alone cannot distinguish between these quite smooth spectral models given the typically bandpass available to X-ray observatories.

Studies of X-ray variability can give complementary information,
potentially breaking this deadlock. The 
X-ray reflection model predicts time delays between variations in the
primary continuum, and the reflected response of the disc (\citealt{Stella90};
\citealt{Reynolds99}).  The primary features of a disc reflected
spectrum include fluorescence/resonance line emission, particularly
from iron, and a curved continuum at high energies.  The
signal-to-noise around the $\sim 6.4$~keV Fe K$\alpha$ line is often rather low,
but if the disc surface is moderately ionised there could also be strong
soft X-ray emission (e.g. \citealt{RossFabian05}), and this soft
emission will also lag behind the illuminating continuum. 
However, these lags are simply the light
travel time to the reflector, so should be short ($10 R_{\rm g}/c\sim
500$~seconds for $M_{\rm BH}=10^7 M_\odot$) so are best 
probed by the fastest variability.  Concentrating on
fast variability is also necessary as accreting black holes (BHs) seem to display hard
lags at low frequencies --- slow variations in the hard energy bands
are delayed with respect to the soft energy bands
(\citealt{miyamoto89}; \citealt{nowak99}a; \citeyear{nowak99b}b; \citealt{vaughan03a}a; \citeyear{vaughan03b}b; \citealt{mchardy04}).  The favoured model
for the origin of hard lags is the radial propagation of accretion
rate fluctuations through a stratified emission region
(e.g. \citealt{kotov01}, \citealt{arevalouttley06}). 
In combination, these two effects predict a switch from hard
(propagation) lags at lower frequencies to soft (reflection) lags at
higher frequencies.  This pattern was tentatively observed in \citet{mchardy07}, and has now been observed in 
$\sim 20$ AGN (e.g. \citealt{fabian09}; \citealt{emmanoulopoulos11}; \citealt{zoghbi11a}; \citealt{alston13b}; \citealt{cackett13}; \citealt{demarco13lags}; \citealt{kara13a}), and
one black hole X-ray binary (XRB; \citealt{uttley2011}).

Narrow Line Seyfert 1 galaxies (NLS1s) are the targets of choice for much of
this work as they are typically brighter in soft X-rays, and have low
black hole mass, so their variability timescales are the shortest
amongst AGN, so they can be studied more easily in single observations
at high signal-to-noise.  A subset of NLS1s (the `complex' NLS1s:
\citealt{Gallo06}) can also show much more dramatic variability than
seen in broad line Seyfert 1s (BLS1s), with deep dips in their X-ray light
curves where their spectra appear dominated by reflection
(\citealt{Fabian04}; 2009). Hence these should give the clearest
detections of reverberation lags, contrasting with the `simple' NLS1,
which show only moderate reflection and variability amplitude
\citealt{Gallo06}.  NLS1s typically show stronger soft 
excesses than broad line AGN (e.g. \citealt{boller96}; \citealt{middleton07}), so the soft
lag signature should be strong in all these systems if the soft excess
is from reflection.  However, detailed spectral decomposition shows
that the majority of the strong soft X-ray emission in NLS1s can be
attributed to direct emission from the accretion disc, itself
extending into the soft X-ray bandpass. There is still a `true' soft
excess (which can again be made from either ionised reflection
or an additional component) in addition to the disc and coronal
components, but it is actually smaller than in broad line objects
(\citealt{Jin2012a}; \citeyear{Jin2012b}; \citealt{done2012a}; \citealt{jinETAL13};
hereafter J13).

J13 used frequency resolved spectroscopy (e.g. \citealt{Revnivtsev99})
as another approach to study the origin of the `true' soft
excess in the `simple' NLS1 \pg. The fastest variability
(timescales $\le 5000$~s) has a spectrum in which the soft
excess is less prominent than in the time averaged spectrum, as
expected if the corona is closest to the black hole so has more rapid
variability than the soft excess and disc emission (J13).  However, little
of this rapid variability of the soft excess is correlated with
the 3-10~keV light curve.  This seems to rule out an ionised reflection
origin for the majority of the soft excess, since ionised
reflection produces simultaneously the soft excess and the Fe K$\alpha$ line and reflection continuum at higher energies (J13).

In this paper we explore the time lags from \pg\
as a function of Fourier frequency and of energy. 


\section{Observations and data reduction}
\label{sect:obs}

\pg was observed by \xmm\ for $\sim 123$~ks in December 2011 (OBS ID:
0675320101).  The timing analysis in this paper uses data from the EPIC-pn camera
\citep{struder01} only, due to its higher throughput and time
resolution.  The pn data were taken in small window (SW) mode, which reduces the impact of event pile-up (\citealt{ballet99}; \citealt{davis01}).  The raw data were processed from Observation Data Files 
following standard procedures using the \xmm\ Science Analysis
System (\textsc{sas} v13.0.0) with a 20 arcsec circular extraction region.  Regions with high background were filtered from the data, but these are very minor, and there was no 
sharp rise in background towards the end of the observation.


\section{Lag as a function of frequency}
\label{sect:lagfreq}

\begin{figure}
\begin{center}
\includegraphics[width=0.4\textwidth]{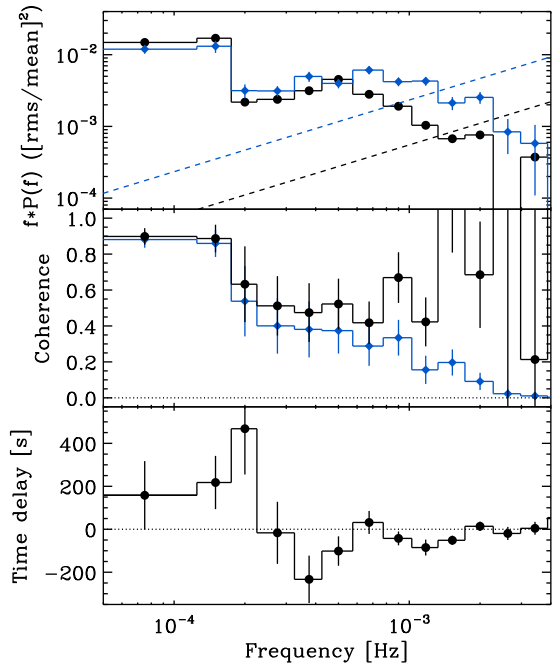}
\caption{Cross-spectral products for the soft (0.3--0.7 keV) and hard
(1.2--4.0 keV) energy bands. Panel (a) shows power spectral density
for the soft (black circles) and hard (blue diamonds) bands. The
dashed lines are the Poisson noise estimates. Panel (b) shows the raw
(blue diamonds) and Poisson noise corrected (black circles) coherence,
see section~\ref{sect:lagfreq} for details. Panel (c) shows the time
lag between the hard and soft band, where a positive values indicate
the hard band lags. }
\label{fig:lag_plot}
\end{center}
\end{figure}

\subsection{The observed lag-frequency}

In this section we explore the cross-spectral products between a soft and hard band.  We use 0.3--0.7~keV as a soft band to maximise the disc and soft
excess components, with little contamination from the hard X-ray
coronal emission, and 1.2-4~keV as a hard band to maximise the coronal
emission, with little contamination from the soft excess or
reflected iron K$\alpha$ emission (J13, see also Fig.~\ref{fig:spec}). These bands
have mean count rates of 5.8 and 0.8 $\rmn{ct\,s}^{-1}$ respectively.

We compute cross spectra following e.g. \citet{vaughannowak97};
\citet{nowak99}; \citet{vaughan03a}.   We estimated cross-spectral
products by first calculating complex cross-spectra values in $M$
non-overlapping segments of time series, and then averaging over the $M$ estimates at each Fourier frequency.  We then averaged in
geometrically spaced frequency bins (each bin spanning a factor $\sim
1.3$ in frequency).  
For the analysis in this paper we use segment
sizes of 20 ks and time bins of 20 s, leaving 6 segments in total.  The choice of segment length and frequency bin averaging was chosen to maximise the number of data points being averaged in the cross-spectrum, whilst still maintaining a high enough frequency resolution to pick out any features in the cross-spectral products.

The upper panel in Fig.~\ref{fig:lag_plot} shows the Poisson noise subtracted power spectra for the hard (blue) and soft (black) bands.  These are very similar to those in J13 and are only given here for completeness.
The Poisson noise level is estimated using equation A2 of \citet{vaughan03a}, and is indicated by the dashed lines. 

From the cross-spectra we get the coherence between the hard and soft band (e.g. \citealt{bendatpiersol86}).  The coherence is defined between [0,1], where 1 is perfect coherence and 0 is perfect incoherence.  This gives an estimate of the linear correlation between the two bands, i.e. how much of the variability in one band can be linearly predicted by the other.
The middle panel of Fig.~\ref{fig:lag_plot} shows the `raw' coherence (blue) and the coherence after Poisson noise correction (black), following \citet{vaughannowak97}.  The noise
corrected coherence is high ($\sim 0.9$) for frequencies up to $\sim
1.5 \times 10^{-4}$ Hz, showing that soft and hard band variations are very well
correlated on timescales longer than a few ks, but then
drops to $\sim 0.5$ between $\sim 1.5-10 \times 10^{-4}$ Hz, and becomes well correlated again at frequencies higher than $\sim 1.5 \times 10^{-3}$ Hz.

From the cross-spectrum we also obtain a phase lag at each frequency,
$\phi(f)$, which we transform into the corresponding time lag $\tau(f)
= \phi(f) / (2 \pi f)$ with errors estimated using raw coherence
(\citealt{vaughannowak97}; \citealt{bendatpiersol86}).  We checked this method produces reliable error estimates when the contribution of Poisson noise is large using Monte Carlo simulations (see e.g. \citealt{alston13b}).  The lower panel shows these frequency dependent time lags between the soft and
hard bands, where we follow convention by using a negative time lag to
indicate the soft band lagging behind the hard band (hereafter `soft
lags').

The lag-frequency spectrum shows a hard lag at frequencies below
$\approx 2 \times 10^{-4}$ Hz, whereas between $\approx 3 \times
10^{-4}$ and $2 \times 10^{-3}$ the soft emission lags the hard, with
a maximum soft lag of $\sim 250$\,s at $\sim 4 \times
10^{-4}$\,Hz. This shape is similar to that seen in other sources, and
seems to be quite common in low-redshift, X-ray variable AGN
(e.g. \citealt{fabian09}; \citealt{emmanoulopoulos11};
\citealt{demarco13lags}).


\subsection{Modelling the lag-frequency}
\label{sect:lag_mod}

\begin{figure}
\begin{center}
\includegraphics[width=0.4\textwidth,angle=0]{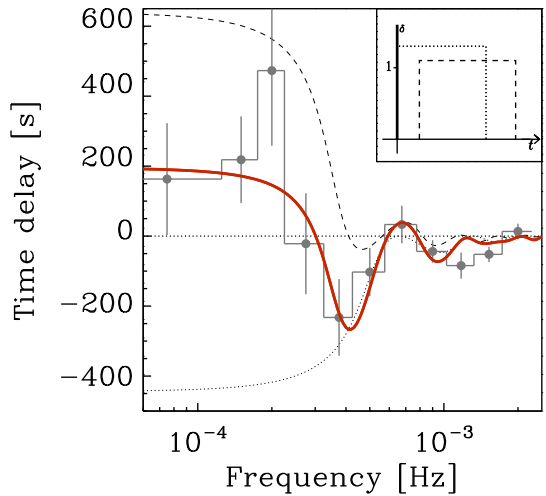}
\caption{The Top hat---Top hat model fit to the lag-frequency spectrum.  The red solid line is the
model fit integrated over the frequency bin, matching the fitting
procedure.  The dashed and dotted lines show the individual transfer functions in each band.  The inset illustrates the top hat and $\delta$-function responses in each band.  See Sec.~\ref{fig:lag_mod} for details.}
\label{fig:lag_mod}
\end{center}
\end{figure}

In this section we model the lag-frequency spectra using simple
analytical models.  Following the approach of \citet{alston13b} (and
references therein) we use a top hat response
functions (or transfer functions when discussing the response function
in the Fourier domain), as well as a power-law dependence on lag-frequency, to model the frequency dependent time delays.
The response functions can be present in either the hard band, the
soft band or both, together with a $\delta$-function response to account
for direct (unreprocessed) emission.  The difference between these two (actually, the complex argument of the product of the two complex transfer functions, see \citealt{alston13b}) predicts the lag-frequency data between the two bands.  We fit the lag-frequency data
using standard $\min(\chi^2)$ fitting techniques, integrating the
model over the bin width, rather than fitting just at the bin
centre. 

A roughly power-law frequency dependence of the hard X-ray time
lag is well established in X-ray binaries (e.g. \citealt{miyamoto88};
\citealt{nowak99}; \citealt{pottschmidt00}), at least at low
frequencies, and consistent results from AGN have been reported
several times (e.g. \citealt{papadakis01}; \citealt{vaughan03a};
\citealt{mchardy04}; \citealt{arevalo06a}).  We therefore consider a
power-law model of the form $\tau(f) = Nf^{-\alpha}$ with $\alpha \approx 1$.
However, a power-law response function alone (with no $\delta$-function)
for the hard band gives an unacceptable fit to the lag-frequency
data ($p < 10^{-8}$), as this predicts only positive (hard)
lags.

We next model the response of the system assuming a $\delta$ function
in the soft band, and a $\delta$ function plus a top hat in the hard
band. 
The $\delta$ functions represent the non-delayed continuum emission in each band\footnote{In the propagating fluctuation model there is an average lag at a given frequency resulting from the energy-dependence of the emissivity function, but each radius contributes both hard and soft band continuum, and so there is some fraction of the emission that is varying near-simultaneously in each band.}.  A $\delta$ function plus top hat is clearly just an approximation to a more complex response function, but is relatively simple to fit and allows for comparison with previous work in this area.
The parameters of the top hat are start time
$t_{0}$, width $w$ and area $S$ (which sets the intensity of delayed
emission relative to the direct emission).  This was used to model
distant reflection by \citet{milleretal10a}, but the physical
picture here is that it models the propagation time delays, as there
is probably little reflection contributing to our `hard' (1.2-4~keV)
bandpass (see \citealt{alston13b}, J13).  However, irrespective of the way
this is interpreted, this can reproduce both positive and negative lags from `ringing' in the Fourier domain as a result of the sharp edges of the
impulse response function \citep{milleretal10a}.  This model also gives a poor fit to the data
($p \sim 10^{-4}$) as the resulting negative lag from ringing 
is too sharp to explain the broad frequency range of the observed
soft lag (see also \citealt{zoghbi11b}; \citealt{emmanoulopoulos11}).

We next considered models including an additional top hat response in the soft band response, where this soft top hat is now physically modelling a reverberation response to the direct emission,
whereas the top hat in the hard band is modelling propagation lags. 
Thus the total model is a $\delta$ function plus top hat in both hard and soft
bands, giving a total of 6 free parameters (3 for each top hat).  We find a
good fit to the data with $\chi^2$ = 9 / 5 d.o.f ($p = 0.1$). The full
resolution model is shown as the red solid line
(Fig.~\ref{fig:lag_mod}). The hard response parameters are: $t_{0}^{h}
= 330\pm80 {\rm s}$, $w^{h} = 1780 \pm 220 {\rm s}$, and $S^{h} = 1.1
\pm 0.3$, while the soft response parameters are: $t_{0}^{s} = 0 \pm
100 {\rm s}$, width $w^{s} = 1520 \pm 140 {\rm s}$, and scaling
fraction $S^{s} = 1.4 \pm 0.5$.  


The lags due to any one component of the combined model can be isolated by computing the lag-frequency curve assuming that the response function in the other band is simply a $\delta$ function.  If
the soft band response is replaced by a $\delta$ function then the top hat
in the hard causes the hard band variations to lag the soft band
variations (dashed line). Conversely if the hard band response is
replaced by a  $\delta$ function then the top hat
in the soft band causes the soft to lag the hard (dotted line), but
the soft lag extends to higher frequencies due to the narrower width
of the top hat in the soft band than in the hard. The lags measured between the two bands, as a function of frequency, is a combination of these lags on each band which may be operating in different directions and so partially cancelling out. The lag from the combined model is approximately the sum of the lag contributions from each model component (these two are not in general exactly equal as the phase difference between two transfer functions is computed after averaging the real and imaginary components of each transfer function over the finite frequency bins).


The typical propagation lags of the hard band behind the soft band are
$\sim 1000$~s, while the typical reverberation timescale lags of the
soft band behind the hard band are of order $\sim 750$~s.  It is this
timescale which gives the size scale of the reprocessor, not the lag
of $\sim 250$~seconds as given from the cross-spectrum as the latter
does not include the effect of dilution of the lagged flux (top hat) by the
direct flux ($\delta$ function) (see e.g. \citealt{milleretal10a};
\citealt{Wilkins12}).  Hence the reprocessing typically occurs on size
scales of $15R_{\rm g}$ for a black hole mass of $10^7$\Msun.

We note that the model still does not quite match the soft lags
observed at the highest frequencies. We could improve the fit by
adding another top hat in the soft band, starting from $t_0^{s2}=0$ with
width of $\sim$~few hundred seconds. This would skew the response
more towards shorter timescales i.e. to weight the reverberation
towards smaller distance material.  Given the limited frequency
resolution, such a complex model will over-fit the data.  We note that improvement in fitting the highest frequency soft lags may also be achieved with the use of `realistic' response function, rather than a simple top hat plus $\delta$ function in the soft band (e.g. \citealt{CampanaStella95}; \citealt{reynolds2000}; \citealt{WilkinsFabian13}).


\section{The frequency resolved lag-energy spectrum}
\label{sect:lagen}

\subsection{The observed lag-energy spectra}

A lag-energy spectrum can be calculated over a given range in
frequency by estimating the cross-spectral lag between a light curve
in each energy band with respect to the frequency resolved light curve
over a broad reference band (e.g. \citealt{zoghbi11b}).  We take our reference
band as the hard band in the previous lag-frequency work i.e. the
1.2--4.0 keV light curve {\em minus} the energy band for which the lag
is being computed so that we never have correlated Poisson noise.  
This reference band is chosen to maximise the lags between the coronal
emission and the soft excess and reflected iron K$\alpha$ emission (see Fig.~1 in J13).  This reduces the contribution to the reference band from the uncorrelated (incoherent) softest energy variations (J13).
A positive lag indicates the given energy bin lags the broad reference
band (so soft lags are now positive).  The lag has not been shifted to
a zero level, so the lag represents the average lag or lead of that
energy band to the reference band.

\begin{figure}
\begin{center}
\includegraphics[width=0.4\textwidth,angle=0]{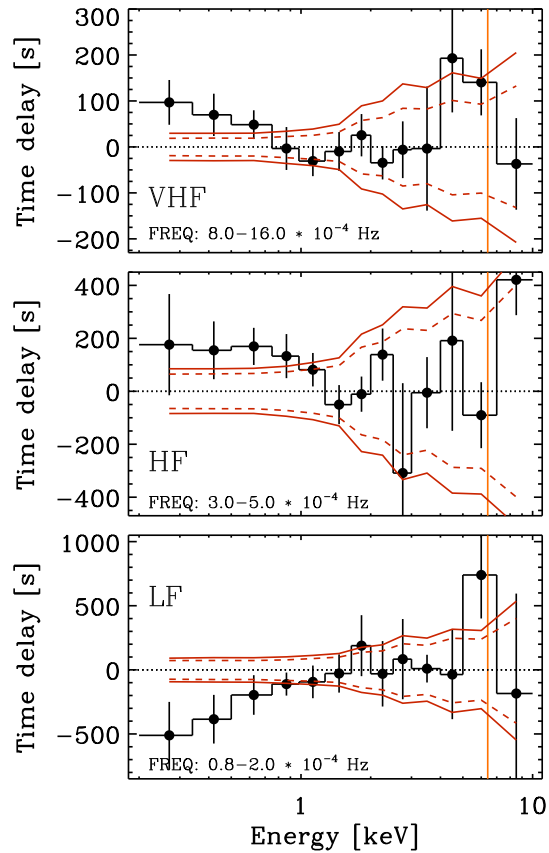}
\caption{Lag-energy spectrum for low (LF; $0.8-2.0 \times 10^{-4}$\,Hz), high (HF; $3-6 \times 10^{-4}$\,Hz) and very high (VHF; $8-14 \times 10^{-4}$\,Hz) frequencies. A 1.2--4.0 keV reference band was used. The red dashed and solid lines are the 90 and 95 per cent confidence intervals respectively, on the assumption of zero `true' lag.}
\label{fig:lag_en_plot}
\end{center}
\end{figure}

\begin{figure}
\begin{center}
\includegraphics[width=0.42\textwidth,angle=0]{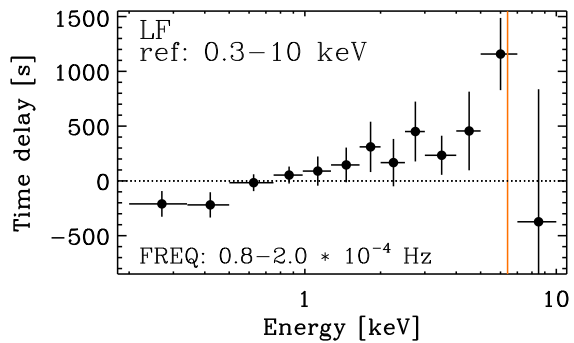}
\caption{Lag-energy spectrum for low frequencies (LF; $0.8-2.0 \times 10^{-4}$\,Hz) using a 0.3--10.0 keV reference band.  The plot shows a continual increase in lag from low to high energies, and are consistent with the lower panel in Fig.~\ref{fig:lag_en_plot} where a 1.2--4.0 keV reference band is used.}
\label{fig:lag_en_LF}
\end{center}
\end{figure}
Motivated by the lag-frequency plot, 
we calculate the lag-energy over three frequency ranges, one focused
on the hard lags ($0.8-2\times 10^{-4}$~Hz, low frequency, hereafter
LF), one centred around the most negative (soft) lags ($3-5\times
10^{-4}$~Hz, high frequency, hereafter HF), and one spanning the
remaining negative (soft) lags ($8-16\times 10^{-4}$~Hz, very high
frequency, hereafter VHF).

Figure~\ref{fig:lag_en_plot} shows the lag as a function of energy for
the LF, HF and VHF
frequencies.  At LF (lower
panel) the soft energies lead the 1.2-4.0~keV reference band, 
with the delay increasing with energy separation.  This is
consistent with results found in BH-XRBs (e.g \citealt{miyamoto89};
\citealt{nowak99}) and AGN (e.g. \citealt{arevaloetal08};
\citealt{zoghbi11b}; \citealt{kara13b}).  A possible ($2\sigma$
significance) lag in the iron K band can be seen, which
lags behind the 1.2--4.0~keV reference band by $\sim 500-1000$\,s.
In Fig.\ref{fig:lag_en_LF} we show the LF lag-energy spectrum computed using the broad 0.3-10 keV energy band.  A continual increase in the lag with energy is seen, and is consistent with the LF lag-energy plot of Fig.~\ref{fig:lag_en_plot}.

At HF (middle panel), the softer energies lag 
the reference band, in
agreement with the lag-frequency spectrum in
Fig~\ref{fig:lag_mod}.  The lag below $\sim 1$\,keV is $\sim
200$\,s behind the reference band.  No iron K$\alpha$
lag can be seen in the data.

At VHF (upper panel in Fig.~\ref{fig:lag_en_plot}) the soft energies
(below $\sim 0.6$\,keV) lag the reference band by up to 100~s.  Also
evident is a feature around iron K$\alpha$ with a lag of $\sim
150$\,s.

The difference in lag-energy between the two soft lag frequency ranges
(HF and VHF) is surprising. This is the first time that the soft lags
have been examined as a function of frequency, and the differences
indicate that different processes may be dominating at these two
different timescales.  We assess the significance of this result
firstly by assessing the significance of the lags in the VHF as this
regime is more susceptible to Poisson noise, particularly at high
energies. We use $10^4$ Monte Carlo simulations of well-correlated light
curves with zero `true lag', with the properties (power spectra, count rates) of the real
data.  The 90 and 95 per cent confidence intervals are shown as the red
dashed and solid lines in Fig.~\ref{fig:lag_en_plot}.  The
low energy continuum (below 0.6~keV) and iron line lags are both
detected at $\sim 2\sigma$ significance.  We also show the 90 and 95
percent confidence intervals for the HF and LF in
Fig.~\ref{fig:lag_en_plot}.  The soft lag is detected at $> 2\sigma$
confidence at energies below 1~keV in the HF.  At LF, the soft lead at
energies below $\sim 0.8$\,keV is detected at $> 2\sigma$, as well as
the lag in the iron K$\alpha$ band.


\subsection{Modelling the lag-energy spectra}

\begin{figure*}
\begin{center}
\centering
\mbox{\subfigure{\includegraphics[width=0.32\textwidth,angle=90]{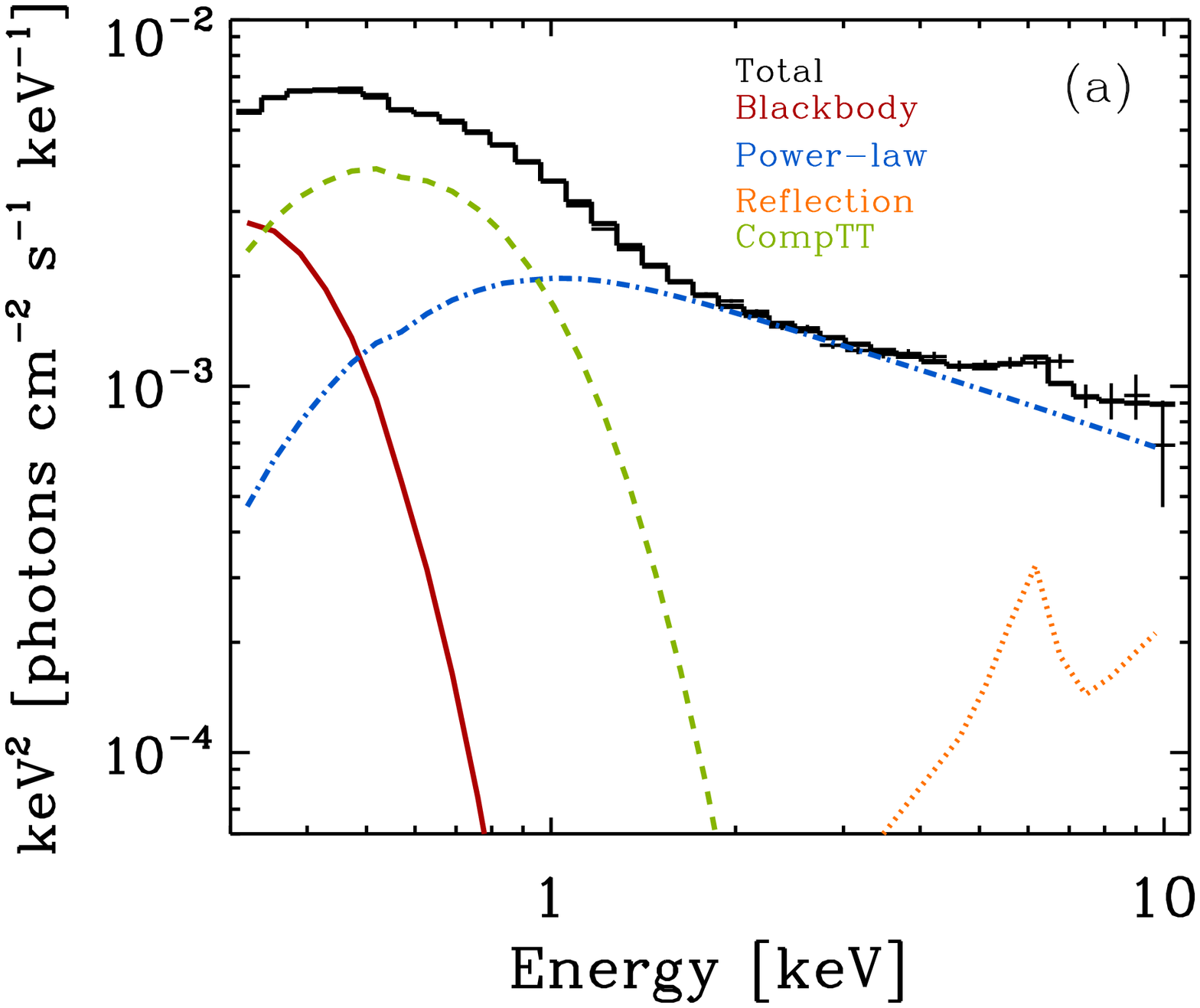}}
\hspace{20 mm}.
\subfigure{\includegraphics[width=0.32\textwidth,angle=90]{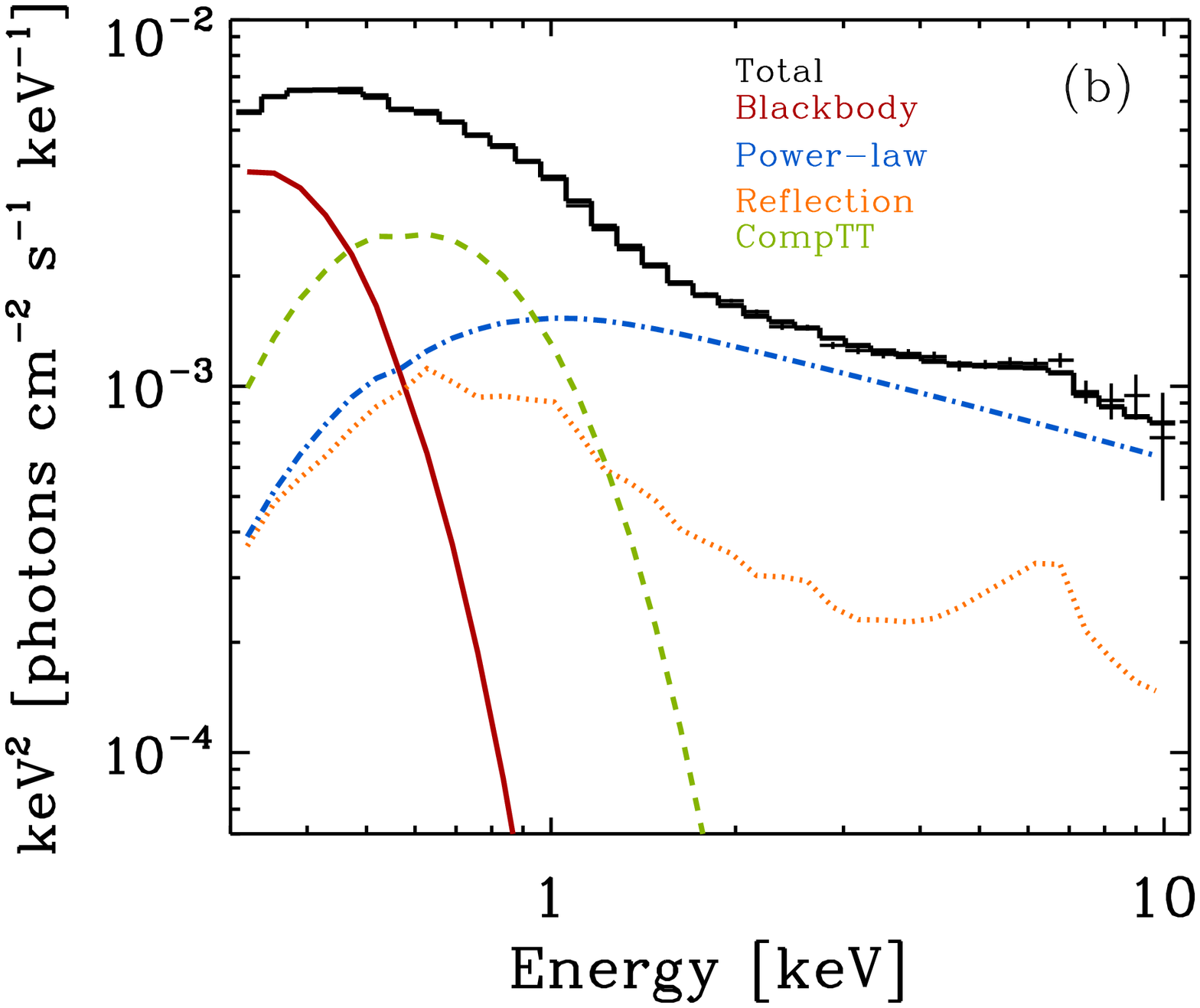}} }
\caption{a) Left: neutral reflection model: blackbody (red solid), an absorbed power-law emission (blue dot-dashed), separate soft excess component (green dashed), and neutral reflection (orange dotted). b) Right: ionised reflection model including a blackbody (red solid), absorbed power-law emission (blue dot-dashed), a separate soft excess component (green dashed), and ionised reflection (orange dotted).  The data are the `unfolded' time-averaged EPIC-pn spectra (see J13).}
\label{fig:spec}
\end{center}
\end{figure*}

\begin{figure*}
\begin{center}
\mbox{\subfigure{\includegraphics[width=0.4\textwidth,angle=0]{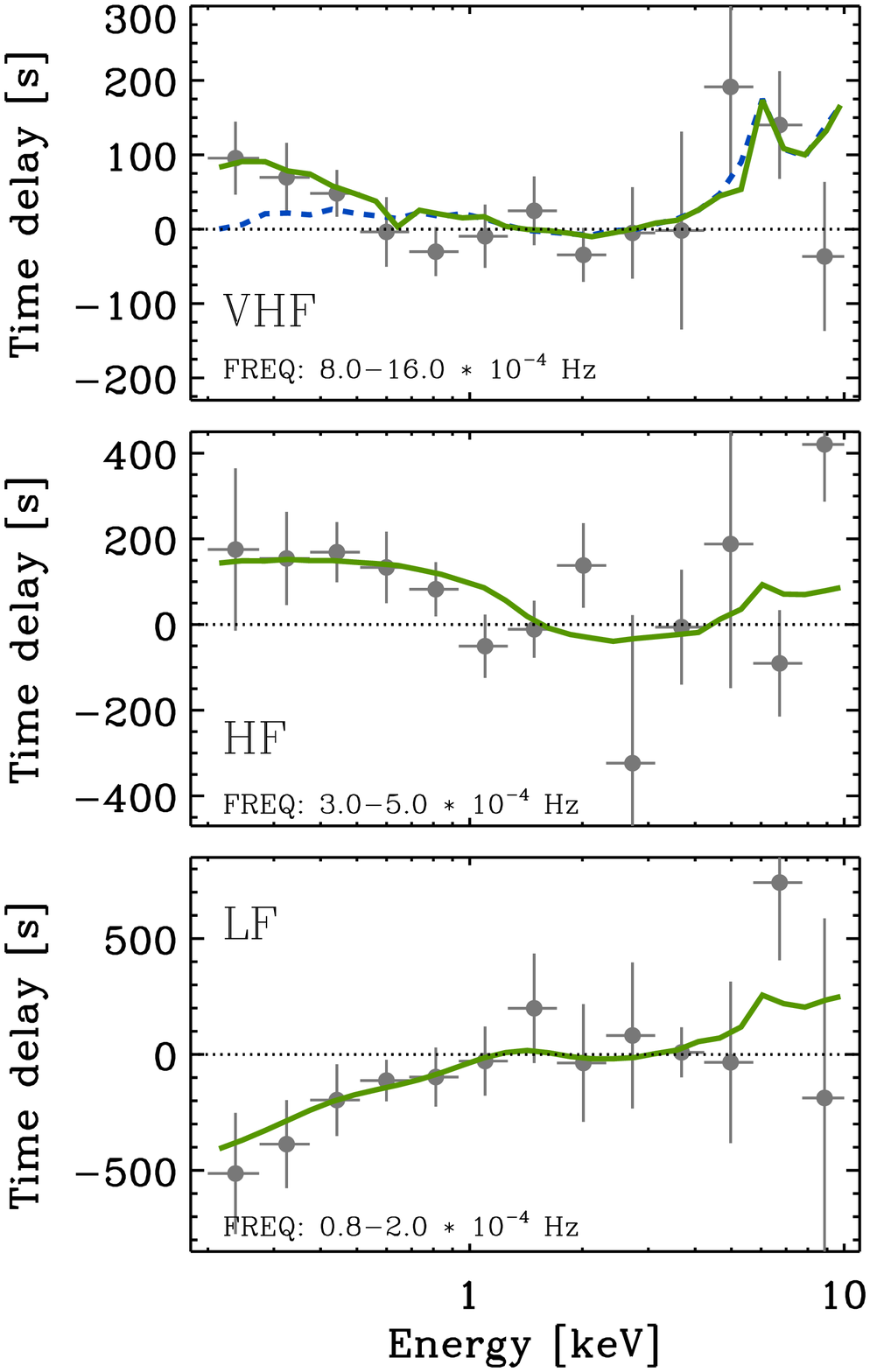}}
\hspace{20 mm}.
\subfigure{\includegraphics[width=0.4\textwidth,angle=0]{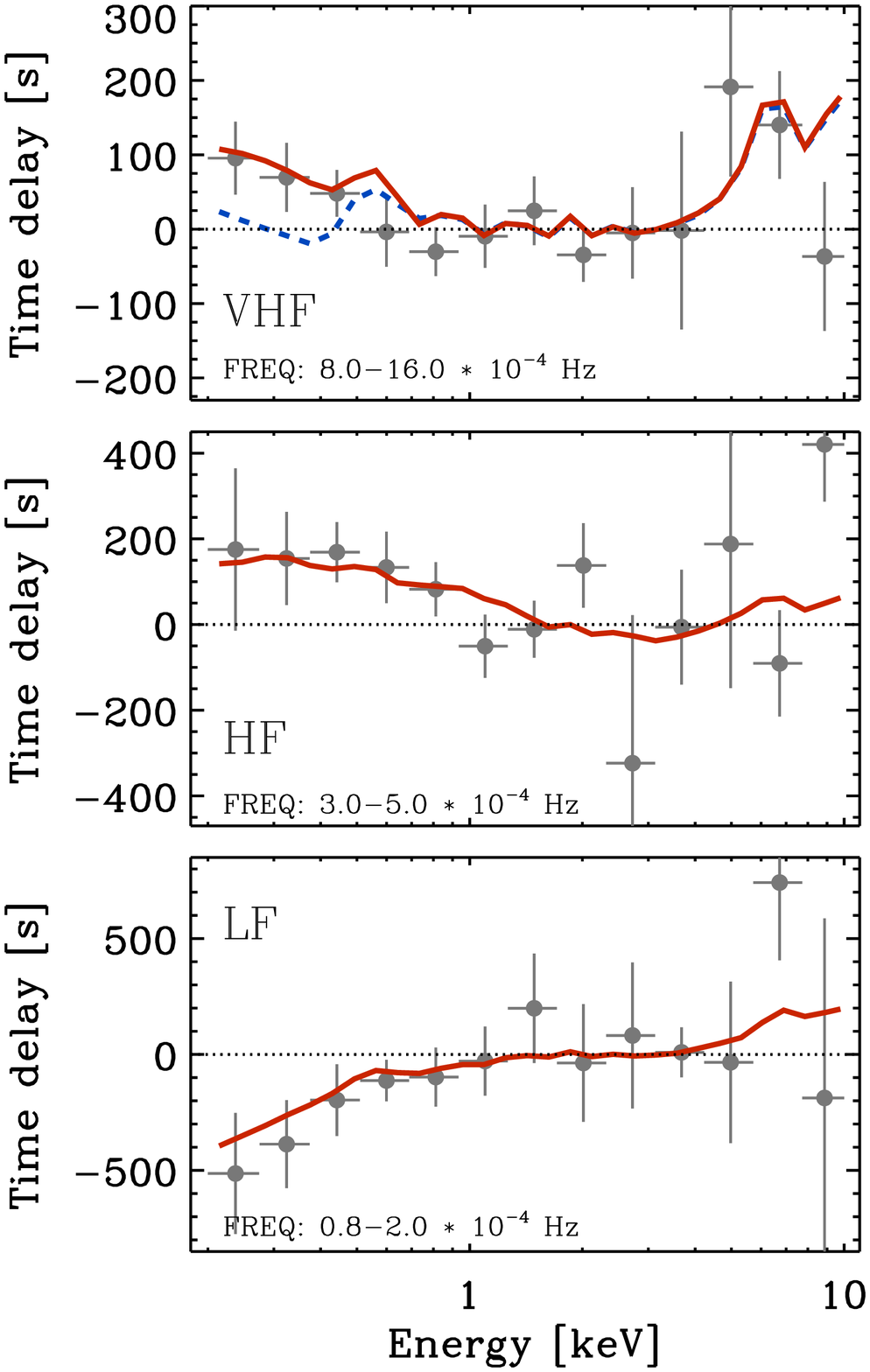}} }
\caption{a) Left: simulated lag-energy spectra for the neutral reflection model, at low (LF; $0.8-2.0 \times 10^{-4}$\,Hz), high (HF; $3-6 \times 10^{-4}$\,Hz) and very high (VHF; $8-14 \times 10^{-4}$\,Hz) frequencies.  A 1.2--4.0 keV reference band was used.  The green line represents the model including the black body, whereas the blue dashed line does not include a lag from this component.
b) Right: simulated lag-energy spectrum for the ionised reflection model, for the LF, HF and VHF ranges. The red line represents the models including the black body, whereas the blue dashed line does not include a lag from this component.}
\label{fig:lagenmod}
\end{center}
\end{figure*}

J13 presented several spectral models to explain the time averaged
spectrum and the covariance spectra.  In this section we re-examine
these models in terms of their lag-energy predictions.  We model the
resulting lag-energy spectra using simulations, similar to the method
of \citet{kara13b}.  We simulated identical sets of light curves for
each energy bin, including independent Poisson noise in each.
The light curves of each component (at each energy) were time-shifted and
then combined weighted by the strength of the component in the mean
spectrum at a given energy.  The lag-energy spectra at each frequency were modelled independently, with each component having a different magnitude at each frequency.  This is clearly a 
simplification of the effects of an impulse response function on each
spectral component,  but
allows us to approximate the shape of the lag-energy spectrum
resulting from the combined effects of multiple components.

\subsubsection{Neutral reflection and a separate soft excess}

We first use the spectral components
identified in J13 (see Fig.~\ref{fig:spec}a), where there is a blackbody from intrinsic disc
emission (red),  a separate soft excess component (green) which
provides seed photons to produce the coronal emission from Compton
upscattering (blue).  This coronal emission illuminates the disc and
produces a (mostly) neutral reflection spectrum (black) from a region 
with inner radius of $\sim 15R_g$.  The soft excess component is modelled with {\sc comptt} (\citealt{TitarchukLyubarskij95}, representing incomplete thermalisation of the inner accretion disc (see \citealt{done2012a}).

For slow variability, where we are seeing propagation time delays, we
assume that the blackbody disc emission leads, followed by the soft
excess, followed by the corona.  This gives a good match to the
LF lag-energy spectra if the soft excess lags 400 seconds
behind the blackbody, and the corona lags by 1500 seconds behind
the blackbody.  A reflection lag of 2~ks behind the corona is included, to model any reflection that is occurring on top of the propagation delays at LF.  The soft band emission then leads the hard band by
$\sim 1000$ seconds, as required by the lag-frequency fits (see
Fig.~\ref{fig:lag_mod}).  These simulated LF lag-energy spectra are shown in the
lower panel of Fig.~\ref{fig:lagenmod}.

The HF lag energy can be matched by assuming that some part of the
soft excess emission is reprocessed on short timescales, so that it lags behind the corona
by 300 seconds (middle panel in Fig.~\ref{fig:lagenmod}).  No iron K$\alpha$ is seen in the data at this frequency, but we include a 300 second lag for this feature.  If an iron K$\alpha$ reflection feature is observed at VHF, and at LF with a longer delay than in the VHF, then we might also expect to see this feature at HF.  Its presence in the LF suggests this feature has not been smeared out over the timescale of interest, and hence why we expect to observe it at HF.

An iron line is instead seen in the VHF lag-energy plot. This can be
well described if some fraction of the reflected
emission lags 200 seconds behind the fastest coronal
variability. However, this cannot reproduce the low energy lag seen in
the VHF (blue simulations in upper panel of Fig.~\ref{fig:lagenmod}).  This is better
fit by the shape of the blackbody disc. The green simulations in
the upper panel of Fig.~\ref{fig:lagenmod} show the VHF lag-energy prediction if
both the blackbody and reflected emission lag 200 seconds behind the
fastest coronal variability.

\subsubsection{Ionized reflection and a separate soft excess}

The model above is an extreme interpretation of the spectrum. The
reflector is likely to be ionized to some extent, so we explore how
this might change the reverberation predictions.  We allow reflection
to be ionised in the fits.  We describe this using the {\sc rfxconv}
model (\citealt{kolehmainen11}), which is a convolution
version of the Ross \& Fabian (2005) ionized reflection models, 
but also include an additional soft
excess as is required by the fast covariance spectra (J13).  However,
this gives a best fitting model ($\chi^2$ = 2204 / 1799 d.o.f) to the time averaged EPIC spectrum 
for fairly low ionisation $\log \xi =1.3$, which is
again not highly smeared by relativistic effects ($R_{\rm in}=10R_{\rm g}$ for
standard emissivity of $\eta(r) \propto r^{-3}$).  This does not produce much more low energy
reflection than the original models, so instead we fix $\log \xi =3$
to explore the impact of the maximal contribution of reflection at low
energies.  This gives a slightly worse fit
with $\chi^2$ = 2220 / 1800 d.o.f, and requires $R_{\rm in}=3R_{\rm g}$ in order to smooth the low
energy atomic features, and so implies high spin, in contrast to the
intrinsic disc component in this source which implies low spin
(\citealt{done2013}).  This spectral model is shown in
Fig.~\ref{fig:spec}b.

The lag prediction from this maximal reflection model are shown in Fig.~\ref{fig:lagenmod}b, 
calculated using the same lags for each component as before.  They 
are not dramatically different to those from neutral reflection, possibly because the ionised reflection is still a small fraction of the soft band flux, relative to the other components. 
The additional blackbody variability is still required in the VHF to match
the soft lag seen in these data as the reflected emission follows the
illuminating spectrum, and there is a downturn in the illuminating
spectrum at low energies due to the seed photon energy for Compton
upscattering being within the observed bandpass (J13).  This downturn
limits the amount of reflection at the lowest energies.

\section{Discussion and Conclusions}
\label{sect:disco}

We show that the X-ray time delays as a function of energy and
Fourier-frequency can be constrained in the `simple' NLS1 PG 1244+026
using a 120\,ks XMM-Newton observation. The lag as a function of frequency
between a hard (1.2--4.0\,keV) and a soft (0.3-1~keV) energy band
shows the now  well established switch from a hard lag at low frequencies to a
soft lag at high frequencies. The
maximum soft lag of $\sim 200$ s is at $\sim 4 \times
10^{-4}$\,Hz but there is also evidence for 
a second negative lag of $\sim 100$\,s
at $\sim 1.2 \times 10^{-3}$\,Hz.  

Modelling the lag-frequency spectrum with simple response functions
gives an acceptable fit if both hard and soft bands contain both
direct emission (modelled using a $\delta$-function response) and a
smoothed, time-delayed response (approximated by a top hat response
function).  We interpret the lagged response in two different ways in
the two different bands, with the hard band lag being a result of
propagation time delays while the soft band lag results from
reprocessing of the primary X-ray spectrum.  The soft band response
function has a maximum time delay of $\sim 1500$\,s.  For a black hole mass of $\sim 10^7 \Mbh$,
this corresponds to reprocessing within $\sim 20 R_{\rm g}$ of the
illuminating source.

We consider the lag as a function of energy in three frequency
ranges. The lag-energy data are consistent with the interpretation above.
The lowest frequency lag-energy spectra shows the long
timescale variability has a lag which increases systematically with
energy (Fig.~\ref{fig:lag_en_plot}). This is expected from
propagating fluctuations through the accretion flow,
where the disc, soft excess and harder X-ray emission from a spatially
extended corona are produced at progressively
smaller radii.  The slow variability originates in the disc,
propagates to the soft excess, then propagates to the corona.  Each
component has its own lag, but the smoothly varying change in the
contribution of each component with energy results in a smooth lag-energy spectrum.  We also find evidence for a disc reflection component responding on this timescale.

On shorter timescales, we would not expect the propagating
fluctuations to correlate at all across the spectrum, as the disc and
soft excess probably cannot vary much intrinsically on these
timescales.  The soft lag at these frequencies is therefore interpreted
as reprocessing of the intrinsic power-law emission.  We split the
frequency band over which the soft lag is seen into two, and
find evidence for a change in the lag-energy spectra with frequency.
The high frequency (HF) variability appears to show a stronger reprocessing response in the
soft excess rather than in the reflected component; there is no
significant lag at iron K$\alpha$ seen in the data, but only at energies below
1.2~keV.  The VHF has a lag at iron (and redward) energies,
but this is accompanied not by the soft excess but by the
blackbody. 

The hard X-ray emission from the corona is consistent with an origin
in inverse-Compton scattering.  The downturn 
below 0.8~keV in the covariance indicates that the source of seed
photons is the soft excess (J13). This strongly limits the amount of
reflection which can contribute to the spectrum below 1~keV.  Thus
reflection alone is not the only source of lags in this object.  An
obvious additional source is thermal reprocessing - the thermalisation of the
non-reflected, absorbed emission, which leads to heating of the disc
and soft excess regions as they respond to increased
illumination. This could give a physical explanation as to why these
components are seen to reverberate along with the reflection
component. However, it is strange that the soft excess emission 
requires typically longer to respond to the  coronal illumination than the
blackbody and iron line.  The propagation lags clearly show that the soft excess region lags the blackbody, and under the assumption that these are modulated by the same inwardly propagating fluctuations, is smaller and closer in than the blackbody.  We would expect its reprocessed signature to 
be more evident at higher frequencies than the blackbody 
as it is closer to the coronal emission region. It may be that 
the region producing the soft excess has little reflected or reprocessed
flux (perhaps because it is too highly ionised), but that it responds to the
corona via propagation lags on the blackbody disc response.  

The hard band PSD (Fig.~\ref{fig:lag_plot}) shows weaker variability power than the soft band at frequencies below $\sim 2 \times 10^{-4}$\,Hz (LF).  At frequencies above  $\sim 2 \times 10^{-4}$\,Hz (HF) the power in each band is comparable, with the hard showing a slight excess in power.  The hard band shows an obvious excess in power exactly where the VHF soft lag occurs.  The coherence also drops to $\sim 0.5$ just above the HF band.  This drop in coherence is most easily explained as due to the `cross over' between two independently varying components with quite different energy spectra.  At LF and VHF bands, where one component dominates the PSD in both bands, the coherence is high since each component is correlated with itself.

Clearly the timing and variability properties of this simple spectrum
source are very interesting.  Better estimates of the lag-energy and
lag-frequency properties, only possible with much longer observations,
are needed to fully understand the causal connection between the emission components
in this highly promising source.


\section*{Additional note}

After the original submission of this manuscript we learnt that the X-ray time lags in PG 1244+026 were simultaneously and independently studied by \citet{kara13sub}.  That submitted paper focuses on the iron K$\alpha$ reverberation lags at low frequencies, whereas this paper has focused on understanding the soft lags, and the time lags observed up to higher frequencies.  The data analysis in the two papers are consistent, with the differences in the lag energy spectra presented in each paper resulting from the choice of reference band.  These two papers highlight the need to better constrain the variability and time delays up to higher frequencies in this source.


\section*{Acknowledgements}

We thank the anonymous referee for the thorough reading of our manuscript.  WNA acknowledges support from an STFC studentship.  This paper is based on observations obtained with \xmm, an ESA science mission with
instruments and contributions directly funded by ESA Member States and the USA (NASA). 

\footnotesize{
\bibliographystyle{mn2e}
\bibliography{pg1244_lags}
}


\label{lastpage}


\end{document}